\documentclass{article}

\usepackage{PRIMEarxiv}

\usepackage[utf8]{inputenc} 
\usepackage[T1]{fontenc}    
\usepackage{hyperref}       
\usepackage{url}            
\usepackage{booktabs}       
\usepackage{amsfonts}       
\usepackage{nicefrac}       
\usepackage{microtype}      
\usepackage{lipsum}
\usepackage{fancyhdr}       
\usepackage{graphicx}       
\graphicspath{{media/}}     
\usepackage{xcolor}
\usepackage{amsmath} 
\usepackage{multirow} 
\usepackage[table]{xcolor}
\definecolor{darkmossgreen}{RGB}{180, 210, 180}
\definecolor{dashedarrow}{HTML}{00B2B2}
\usepackage{booktabs}
\usepackage{makecell}
\usepackage{tabularx}
\usepackage{adjustbox}
\usepackage{caption}
\usepackage{ragged2e}
\usepackage{enumitem}
\usepackage{threeparttable}
\title{Median2Median: 
Zero-shot Suppression of Structured Noise in Images}

\author{
\begin{tabular}{c c}
Jianxu Wang\textsuperscript{1} & Ge Wang\textsuperscript{1*} \\[2pt]
\texttt{wangj68@rpi.edu} & \texttt{wangg6@rpi.edu} \\
\end{tabular} \\[14pt]  
\textsuperscript{1}\,Biomedical Imaging Center, Center for Computational Innovations,\\[1pt]
Center for Biotechnology \& Interdisciplinary Studies, Department of Biomedical Engineering,\\[1pt]
School of Engineering, Rensselaer Polytechnic Institute, Troy, New York 12180, USA\\[6pt]
\textsuperscript{*}\,Corresponding author
}

\begin{document}
\maketitle

\begin{abstract}
Image denoising is a fundamental problem in computer vision and medical imaging. However, real-world images are often degraded by structured noise with strong anisotropic correlations that existing methods struggle to remove. Most data-driven approaches rely on large datasets with high-quality labels and still suffer from limited generalizability, whereas existing zero-shot methods avoid this limitation but remain effective only for independent and identically distributed (i.i.d.) noise. To address this gap, we propose Median2Median (M2M), a zero-shot denoising framework designed for structured noise. M2M introduces a novel sampling strategy that generates pseudo-independent sub-image pairs from a single noisy input. This strategy leverages directional interpolation and generalized median filtering to adaptively exclude values distorted by structured artifacts. To further enlarge the effective sampling space and eliminate systematic bias, a randomized assignment strategy is employed, ensuring that the sampled sub-image pairs are suitable for Noise2Noise training. In our realistic simulation studies, M2M performs on par with state-of-the-art zero-shot methods under i.i.d. noise, while consistently outperforming them under correlated noise. These findings establish M2M as an efficient, data-free solution for structured noise suppression and mark the first step toward effective zero-shot denoising beyond the strict i.i.d. assumption.
\end{abstract}


\section{Introduction}
Image noise primarily arises during data acquisition, stemming from sensor electronics, photon statistics, and variations in imaging conditions such as illumination and scene dynamics. Additional artifacts may also be introduced during transmission and compression. The resulting degradations can be stochastic (e.g., Poisson or Gaussian) or structured (e.g., speckle or banding), both of which reduce the signal-to-noise ratio, impair visual quality, and hinder downstream analysis. Denoising methods are therefore needed to recover the underlying clean image, either by modeling the corruption process or by exploiting image priors such as self-similarity and sparsity~\cite{goyal2020image}. Across biomedical imaging modalities such as computed tomography (CT), nuclear imaging (PET and SPECT), ultrasound imaging, and fluorescence microscopy, as well as natural images captured under challenging conditions, denoising remains a foundational task before further processing and analysis.

Conventional denoising methods typically assume that noise follows a known statistical model, such as a Gaussian distribution or a mixture of well-characterized distributions~\cite{dabov2007image, moulin1999analysis}. These assumptions allow classical methods to perform noise removal in a principled manner. However, most real-world noise is considerably more complex, being neither independent nor accurately represented by analytic models.

In recent years, deep learning-based denoising methods have emerged with impressive successes. Early supervised methods train neural networks to learn end-to-end mappings from noisy observations to clean images~\cite{zhang2017beyond, yang2018low}. The Noise2Noise (N2N) framework~\cite{lehtinen2018noise2noise} relaxes this requirement by showing that supervised denoising can be performed using pairs of noisy observations of the same clean image, each corrupted by independent and identically distributed (i.d.d.) zero-mean noise. However, this training framework still requires large domain-specific datasets that must be carefully collected and curated, which is often costly and time-consuming. Moreover, such data-driven models typically generalize poorly to unseen data when the test distribution differs from that of the training set.

\begin{figure}[t]
\centering
\includegraphics[width=\linewidth]{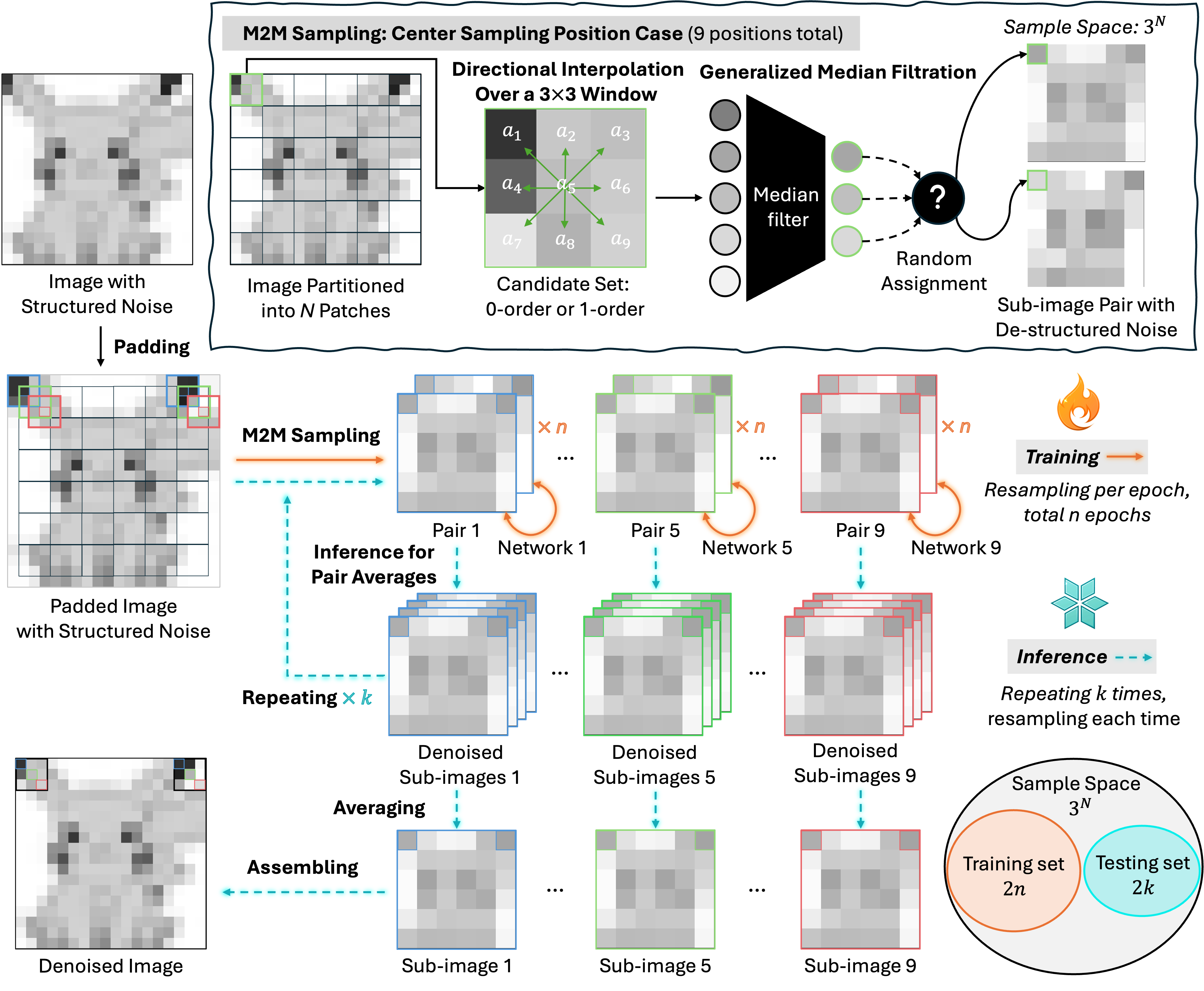}
\caption{Overview of the proposed M2M framework. 
The top box illustrates M2M sampling at the center position in the first patch, where directional interpolation and median filtering are applied to obtain de-structured pixel values from an image corrupted by structured noise. Below the top box are the training and inference pipelines, where the \textcolor{orange}{solid arrows} indicate the training process and the \textcolor{dashedarrow}{dashed arrows} denote the inference process. The bottom right panel shows the sampling space ($3^N$), the training set ($2n$), and the testing set ($2k$).}
\label{fig:overview}
\end{figure}

In contrast to supervised methods, zero-shot denoising offers a compelling alternative by training a network directly on a single noisy input, without requiring any external dataset. Noise2Void~\cite{krull2019noise2void} and Noise2Self~\cite{batson2019noise2self} adopt blind-spot networks that exclude the central pixel during training, so the model learns to infer the underlying signal from contextual information rather than fitting the noise. Although effective in theory and practice, these methods are computationally intensive and thus less feasible for real-world applications. More recently, inspired by Neighbor2Neighbor~\cite{huang2021neighbor2neighbor}, Noise2Fast~\cite{lequyer2022fast} introduced checkerboard downsampling to efficiently construct noisy pairs, enabling denoising within tens of seconds. 

Subject to the assumption of the N2N-style training framework on the properties of noisy pairs, the aforementioned zero-shot methods fail in the presence of structured noise. This has been a major open problem, since structured noise or weak artifacts are ubiquitous in real-world images across diverse domains, including biomedical imaging, remote sensing, and industrial inspection. Unlike i.d.d. noise, these structured corruptions exhibit correlations along dominant orientations and patterns that are coherent spatially, temporally, or otherwise. As a result, they are challenging to suppress without introducing blur and residual artifacts.

To robustly and cost-effectively suppress structured noise in a zero-shot setting, here we propose Median2Median (M2M), a blind denoising method that leverages characteristics of correlated noise via generalized median filtering to generate pseudo-independent noisy image pairs adaptively from a single input, as illustrated in Fig.~\ref{fig:overview}. The core idea of our method lies in the M2M block-wise sampling strategy. Specifically, a noisy image is first partitioned into non-overlapping $3 \times 3$ patches, each containing nine predefined sampling positions: top-left, top, top-right, left, center, right, bottom-left, bottom, and bottom-right. As an illustration, the top box in Fig.~\ref{fig:overview} shows an example of M2M sampling at the center sampling position in the first patch. For each sampling position, pixel values at the same relative location across all patches are estimated using zero-order or first-order directional interpolation within the corresponding $3 \times 3$ interpolation window. In this process, zero-order interpolation derives the estimate directly from immediate neighbors, whereas first-order interpolation predicts it by averaging symmetric neighbors along predefined orientations. The resulting estimates, together with the pixel value at the target location, constitute a candidate set that may be biased by structured noise while at the same time containing values affected by independent noise. To extract the latter, a median-based filtration is applied, under the assumption that outliers caused by structured noise tend to deviate from the median further. Based on this heuristic consideration, we retain the three values that are closest to the median as representative candidates. To avoid systematic bias introduced by sorting, we use a random assignment strategy (RAS), in which two of these candidates are randomly selected and paired as pixel values of the current sampling position. Repeating the above sampling process across all patches yields paired pseudo-independent noisy pairs that serve as source–target pairs for Noise2Noise training. The randomization introduced by RAS enlarges the effective sampling space and increases diversity. After applying one M2M sampling operation to all nine sampling positions, a total of nine sub-image pairs are obtained, which are distinct and complementary.

During training, new sub-image pairs are resampled for each epoch, and nine independent lightweight networks are trained in parallel, each dedicated to one of the predefined sampling positions. For inference, paired samples from the nine positions are processed by their respective networks, and this procedure is repeated $k$ times. Then, the results are averaged to ensure convergence to the ground truth. The denoised outputs associated with the nine positions are finally reassembled to form the final image. As the entire procedure does not require external data and is adaptive to directional noise structures, M2M exhibits robust generalization and remains effective in the presence of any directionally correlated noise.

Our two main contributions are as follows:
\begin{itemize}[topsep=0.6em, itemsep=0.6em, parsep=0em]
    \item \textbf{Construction of pseudo-independent noisy image pairs from an image corrupted by structured noise.} Compared to existing sampling schemes such as those in N2F, M2M leverages zero-order and first-order directional interpolation to generate multiple signal estimates and applies generalized median filteration to exclude outliers distorted by structured noise, thereby obtaining surrogate samples primarily affected by independent noise. These sampling results can satisfy the assumption of Noise2Noise learning.
    \item \textbf{First zero-shot denoising method that is effective in suppressing structured noise.} M2M is specifically designed to suppress structured or correlated noise, which has been challenging to existing zero-shot denoising methods. Our realistic simulation experiments show that M2M consistently outperforms state-of-the-art zero-shot denoising methods. Up to date, M2M is the only zero-shot technique shown to effectively eliminate correlated noise patterns in an unsupervised setting.
\end{itemize}

\section{Related Work}
\label{sec:related work}
\textbf{Supervised denoising methods} use paired datasets to train neural networks to recover clean images from noisy inputs. Early methods such as DnCNN~\cite{zhang2017beyond} and FFDNet~\cite{zhang2018ffdnet} use clean–noisy image pairs to learn a denoising mapping. This Noise2Clean (N2C) strategy offers a strong performance benchmark and makes no assumption about the underlying noise characteristics, whether the noise is structured or unstructured, following any statistical distribution. In the supervised learning mode, the network is guided by training data to suppress whatever noise in the dataset. However, acquiring clean–noisy image pairs is often labor-intensive, expensive, and even infeasible in practice, particularly in medical imaging tasks. 

Noise2Noise (N2N)~\cite{lehtinen2018noise2noise} introduced a new training framework that learns from pairs of noisy observations of the same underlying image, under the assumption that the noise is zero-mean and i.i.d. This relaxation removes the need for clean ground truth while achieving performance comparable to N2C. However, the requirement of obtaining two acquisitions with i.i.d. noise remains a serious hurdle, especially when repeated imaging is costly, risky, or unstable due to misregistration of subsequent images.

\textbf{Self-supervised denoising methods} remove the need for paired images. Noise2Void~\cite{krull2019noise2void} masks target pixels and trains the network to predict them using only the surrounding context. Similarly, Neighbor2Neighbor~\cite{huang2021neighbor2neighbor} creates pseudo pairs from adjacent pixels within the same image. Despite these innovations, self-supervised denoising still depends on access to large datasets of noisy images. These data-demanding methods, including both supervised and self-supervised approaches, often fail to generalize well. That is, they may degrade in performance when test data deviates from training data distribution-wise.

\textbf{Zero-shot denoising methods} learn to denoise directly from a single noisy image without requiring any external dataset, making them particularly advantageous in data-limited settings. Unlike supervised and self-supervised methods, zero-shot methods are not sensitive to any distribution shift of test data, allowing superior generalizability across noise types and image domains, since training and inference occur on the same image. 

BM3D\cite{dabov2007image} improves image quality by grouping similar patches into 3D blocks, applying collaborative filtration via a transform domain shrinkage and aggregating overlapping estimates to suppress noise while preserving fine details. However, BM3D is by design for Gaussian noise and requires the user to input an estimate of the noise standard deviation, a parameter to which BM3D is highly sensitive. Deep Image Prior (DIP)\cite{ulyanov2018deep} leverages the induction bias of a randomly initialized convolutional neural network as an implicit image prior. It enables denoising without external data by fitting the network directly to the noisy image. However, DIP's performance is closely related to the number of iterations: too few iterations may lead to underfitting to features, while too many iterations risk overfitting to noise. Both BM3D and DIP require user intervention and are therefore regarded as knowledge-based or non-blind zero-shot methods. Their high parameter sensitivity further limits their practicality.

In contrast, blind zero-shot methods operate without requiring any prior knowledge. Noise2Self~\cite{batson2019noise2self} extends masking-based learning into a zero-shot setting by predicting masked pixels using the unmasked context in the same image. Self2Self~\cite{quan2020self2self} denoises a single noisy image by training on stochastic masked versions and averaging multiple stochastic predictions. While its performance rivals that of the supervised methods, Self2Self introduces significant computational overhead due to repeated inferences. To address run-time concerns, Noise2Fast~\cite{lequyer2022fast} accelerates self-supervised learning by constructing training pairs using checkerboard downsampling from a single noisy input. This enables zero-shot denoising in real-time without any prior knowledge of the noise distribution. Zero-shot Noise2Noise (ZS-N2N)~\cite{mansour2023zero} further simplifies sampling by efficiently computing $2{\times}2$ diagonal means through convolution, and introduces a consistency loss to enhance network performance.

\begin{table}[t]
\captionsetup{justification=raggedright,singlelinecheck=false}
\caption{Comparison of self-supervised and zero-shot denoising methods.}
\renewcommand{\arraystretch}{1.3}
\rowcolors{2}{gray!15}{white}
\begin{tabularx}{\textwidth}{>{\raggedright\arraybackslash}p{3cm} 
                                >{\raggedright\arraybackslash}p{3.2cm} 
                                >{\raggedright\arraybackslash}X 
                                >{\raggedright\arraybackslash}p{1.5cm}}
\Xhline{1pt}   
\rowcolor{gray!60}
\textbf{\color{white} Method} & \textbf{\color{white} External Data} & \textbf{\color{white} Characteristics} & \textbf{\color{white} Efficiency} \\
\Xhline{0.8pt} 
Deep Image Prior (2018) & No (single-image) & Randomly initialized CNN directly fits a single noisy image, exploiting the CNN induction bias & \textbf{Slow} \\
Noise2Void (2019) & Usually (dataset, single-image optional) & Blind-spot masking: hide center pixel, predict from neighbors & \textbf{Medium} \\
Noise2Self (2019) & Usually (dataset, single-image optional) & J-invariant masking: randomly mask pixels, predict from remaining ones & \textbf{Medium} \\
Self2Self (2020) & No (single-image) & Dropout augmentation: use random dropout masks for training, and averages multiple dropout predictions at inference & \textbf{Slow} \\
Neighbor2Neighbor (2021) & Yes (dataset) & Neighbor downsampling: randomly split neighboring pixels into two sub-images & \textbf{Medium}\\
Noise2Fast (2021) & No (single-image) & Checkerboard downsampling: form sub-images for training & \textbf{Fast} \\
Zero-Shot Noise2Noise (2023) & No (single-image) & $2{\times}2$ diagonal means + symmetric \& consistency loss & \textbf{Fast} \\
\Xhline{1pt}   
\end{tabularx}
\label{tab:denoising_methods}
\end{table}

Table~\ref{tab:denoising_methods} summarizes the characteristics of the above self-supervised and zero-shot denoising methods. All of these approaches are built on the assumption of i.i.d. noise, and thus remain effective only in such cases while failing under structured noise. However, in real-world applications such as medical imaging, images often exhibit correlated noise or subtle structured artifacts. Addressing this limitation is the central focus of our work.

\section{Methodology}

The effectiveness of Median2Median (M2M) in suppressing structured noise stems from its novel sampling strategy, which integrates directional interpolation with generalized median filtration to generate training pairs corrupted by de-structured noise. Our sampling strategy departs from the existing blind zero-shot denoising methods: rather than relying on random masking or passive spatial subsampling to select neighboring pixels, M2M actively estimates the value of target pixels by adaptively excluding those corrupted by structured artifacts. This procedure allows the construction of pseudo-independent noisy pairs that satisfy the statistical assumption for Noise2Noise learning, even in the presence of highly structured noise. In this section, we describe our M2M approach with key details, including its sampling pattern, filtering operation, network architecture, training technique, and inference scheme.

\subsection{Sampling Pattern}
\label{sec:sampling algorithm}
\subsubsection{Directional Interpolation}
\label{sec:directional interpolation}
\begin{figure}[t]
\centering
\includegraphics[width=\textwidth]{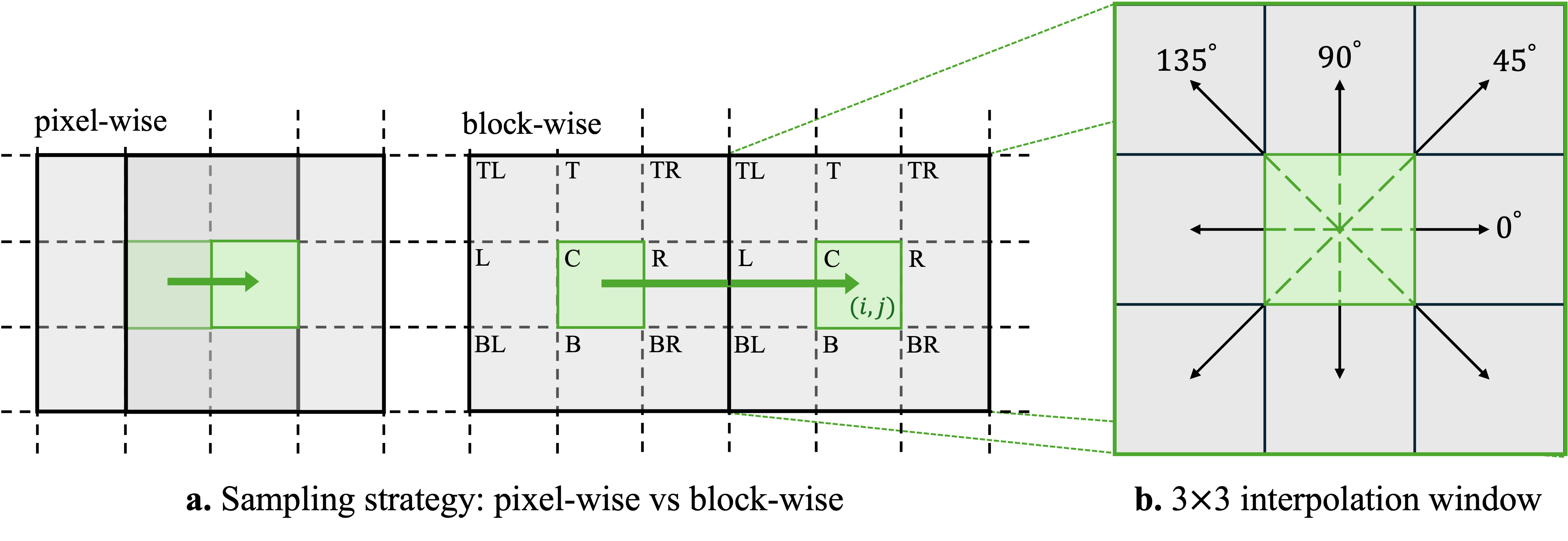}
\caption{Sampling and interpolation processes. 
(a) Comparison of pixel-wise and block-wise sampling strategies: pixel-wise sampling moves a $3 \times 3$ window with a stride $1$ thus a resultant overlap, whereas block-wise sampling partitions the image into non-overlapping $3 \times 3$ patches, each with nine predefined sampling positions—top-left (TL), top (T), top-right (TR), left (L), center (C), right (R), bottom-left (BL), bottom (B), and bottom-right (BR).
(b) A $3 \times 3$ interpolation window used for directional interpolation, illustrated with the center sampling position as an example. Zero-order interpolation directly takes values from neighboring pixels, while first-order interpolation averages symmetric neighbors along each of the four predefined directions ($0^\circ$, $45^\circ$, $90^\circ$, and $135^\circ$).}
\label{fig:blockwise_sampling}
\end{figure}

We begin by introducing the directional interpolation method for extracting reliable signal estimates from a structured noisy environment. Let $I \in \mathbb{R}^{H \times W}$ be a 2D noisy image. To enable uniform partitioning into non-overlapping $3 \times 3$ petches, the image is first extended by reflective padding, which mirrors the boundary pixels. The padded image is denoted by $\tilde{I} \in \mathbb{R}^{\tilde{H} \times \tilde{W}}$, where
\begin{equation}
\tilde{H} = \left\lceil \frac{H}{3} \right\rceil \times 3, \quad \tilde{W} = \left\lceil \frac{W}{3} \right\rceil \times 3.
\end{equation}
In this setting, each $3 \times 3$ patch is associated with nine relative pixel locations---top-left, top, top-right, left, center, right, bottom-left, bottom, and bottom-right---which we define as sampling positions, as shown in Fig.~\ref{fig:blockwise_sampling}a. During sampling, all patches are evaluated at the same relative position to ensure consistency across the image. For each sampling position, a $3 \times 3$ interpolation window is centered on that location, and the corresponding pixel estimate is computed from its neighboring pixel values. By construction, this block-wise design avoids overlap between sampling domains and thus mitigates correlations that would otherwise arise from repeatedly sampling the same pixel. We refer to this as block-wise sampling, in contrast to the more intuitive pixel-wise sampling.

Fig.~\ref{fig:blockwise_sampling}b illustrates the process of performing directional interpolation with a $3 \times 3$ window at the center sampling position of a patch, corresponding to the pixel at coordinate $(i, j)$ in $\tilde{I}$. The interpolation methods considered in M2M include zero-order interpolation using nearest neighbors and first-order interpolation using directional averages.

\textbf{Zero-order directional interpolation.} Zero-order estimates are formed by directly adopting the intensity values of neighboring pixels at the sampling position. Depending on the neighborhood definition, these estimates can be drawn from four axis-aligned or eight surrounding neighbors. Formally, a zero-order estimate with offset $(\Delta i, \Delta j)$ is defined as
\begin{equation}
\label{eq:0order}
\hat{I}_{(\Delta i, \Delta j)}(i, j) = \tilde{I}(i + \Delta i, j + \Delta j),
\end{equation}
where $(\Delta i, \Delta j) \in \mathcal{N}_k$, and $k \in \{4,8\}$ specifies the neighborhood type. The corresponding neighborhood sets are defined as
\begin{equation}
\mathcal{N}_4 = \{(\pm 1, 0), (0, \pm 1)\}, \quad 
\mathcal{N}_8 = \mathcal{N}_4 \cup \{(\pm 1, \pm 1)\}.
\end{equation}

\textbf{First-order directional interpolation.} In contrast, the first-order scheme derives orientation-specific estimates at the sampling position $(i, j)$ by averaging the values of its two symmetric neighbors along four predefined orientations: horizontal ($0^\circ$), vertical ($90^\circ$), diagonal ($45^\circ$), and anti-diagonal ($135^\circ$). Formally, a first-order estimate along a direction $\theta$ is defined as
\begin{equation}
\label{eq:1order}
\hat{I}_\theta(i, j) = \frac{1}{2} \left[ \tilde{I}(i + \Delta i_\theta, j + \Delta j_\theta) + \tilde{I}(i - \Delta i_\theta, j - \Delta j_\theta) \right],
\end{equation}
where $(\Delta i_\theta, \Delta j_\theta)$ is the directional offset for $\theta \in \{0^\circ, 45^\circ, 90^\circ, 135^\circ\}$, corresponding to $(0,1)$, $(-1,1)$, $(1,0)$, and $(-1,-1)$ respectively.

Eqs.~\ref{eq:0order} and \ref{eq:1order} yield two types of interpolation-based estimates for the pixel value at the sampling position $\tilde{I}(i, j)$ within the $3 \times 3$ neighborhood. These estimates primarily capture the noise characteristics along different orientations. Since $\tilde{I}(i, j)$ inherently carries information of the underlying signal at that location, it is also retained as a valid candidate in both cases. Accordingly, the complete candidate sets are defined as
\begin{equation}
\mathcal{S}_0(i, j) = \left\{ \tilde{I}(i, j) \right\} \cup \left\{ \hat{I}_{(\Delta i, \Delta j)}(i, j) \mid (\Delta i, \Delta j) \in \mathcal{N}_k \right\}, \quad k \in \{4, 8\},
\end{equation}
for the zero-order interpolation, and
\begin{equation}
\mathcal{S}_1(i, j) = \left\{ \tilde{I}(i, j) \right\} \cup \left\{ \hat{I}_\theta(i, j) \mid \theta \in \{0^\circ, 45^\circ, 90^\circ, 135^\circ\} \right\},
\end{equation}
for the first-order interpolation. The two sets $\mathcal{S}_0$ and $\mathcal{S}_1$ provide two alternative candidate groups for subsequent filtering and estimation.

\begin{figure}[t]
\centering
\includegraphics[width=\linewidth]{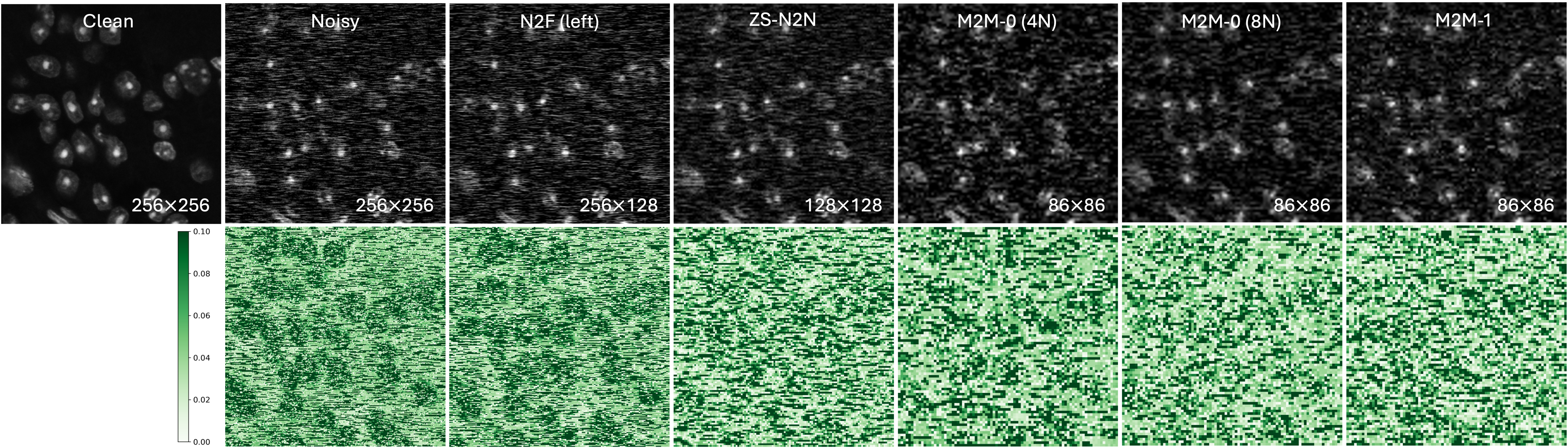}
\caption{Sampling results from a fluorescence microscopic image corrupted by horizontally correlated noise. From left to right: the clean reference, the noisy input, sampling with N2F using checkerboard sampling followed by a squeeze-left operation, sampling with ZS-N2N using $2 \times 2$ diagonal averaging, and sampling with M2M using zero-order interpolation with four-neighborhood, zero-order interpolation with eight-neighborhood, and first-order interpolation. The second row shows the residuals relative to their clean counterparts. It can be observed that N2F and ZS-N2N retain strong horizontal structures inherited from the correlated noise in their sampling results, whereas M2M sampling effectively suppresses such directional interference.}
\label{fig:sampling_results}
\end{figure}

Unlike isotropic i.i.d. noise, which is directionally independent, structured noise is anisotropic so that the estimation accuracy varies with the orientation used for interpolation, often amplifying or suppressing pixel intensities more along a direction of stronger noise correlation. Suppose that a region contains correlated noise along a direction $\theta_s$. In such a case, if we were to apply Neighbor2Neighbor-style sampling as in Noise2Fast \cite{lequyer2022fast} or ZS-N2N \cite{mansour2023zero}, such directional artifacts would persist in the sampling results, as illustrated in Fig.~\ref{fig:sampling_results}. In contrast, the directional interpolation adopted in M2M provides multiple alternative estimates along different orientations. Although estimates along direction $\theta_s$—either the first-order estimate $\hat{I}_{\theta_s}(i, j)$ or the zero-order estimates from neighbors aligned with $\theta_s$—may be noticeably biased, appearing significantly higher or lower than estimates from other directions, by coupling with the generalized median filtration to be introduced next, our method allows for the identification and suppression of biased estimates caused by structured noise.

\subsubsection{Generalized Median Filtration}

Through directional interpolation, each $3 \times 3$ patch in $\tilde{I}$ provides $n$ candidate estimates for a given sampling position, where $n$ depends on the interpolation scheme (e.g., $n=5$ for first-order interpolation or zero-order interpolation with a 4-neighborhood, and $n=9$ for zero-order interpolation with an 8-neighborhood). The values in the candidate set are sorted in ascending order and denoted as $v_{(1)} \leq v_{(2)} \leq \cdots \leq v_{(n)}$. As discussed in Subsection~\ref{sec:directional interpolation}, when an image is corrupted with directionally correlated noise, not all directional estimates are equally reliable. When the interpolation orientation coincides with the direction of structured noise, the corresponding estimate is often biased and deviates significantly from the others.

To suppress the influence of outliers, we apply a generalized median filtering step to each candidate set. Specifically, from the $n$ sorted candidate values, we retain the three values closest to the median and denote the filtered candidate subset as
\begin{equation}
\mathcal{S}_f(i,j) = \{v_{(\lceil n/2 \rceil -1)},\ v_{(\lceil n/2 \rceil)},\ v_{(\lceil n/2 \rceil +1)}\}.
\end{equation}
These median-centered values are considered more reliable, whereas the extremes are discarded because they are more susceptible to bias from structured noise. In other words, while the median filtering method is an established classic denoising method, our generalized median filtering method is intended to construct a pair of noisy patches more favorably, i.e., corrupted with pseudo-independent yet weakened noise.

\subsubsection{Random Assignment}

To enlarge the effective sampling space, we adopt a randomized assignment strategy instead of a fixed scheme. From the filtered subset, two distinct elements are uniformly sampled without replacement to form a de-structured noisy pair. The sampled values, denoted by $m_1$ and $m_2$ with $m_1 \leq m_2$, are drawn as
\begin{equation}
(m_1, m_2) \sim \mathcal{U}_2\!\left(\mathcal{S}_f(i,j)\right),
\label{eq:random1}
\end{equation}
where $\mathcal{U}_2(\cdot)$ denotes uniform sampling of two distinct elements from the given set. This center-aware filtering process suppresses biased directional estimates and yields two surrogate observations that can be regarded as approximately independent realizations of the same underlying signal.

Directly assigning the selected median values $m_1$ and $m_2$ to the sampled images $x_{1}$ and $x_{2}$ introduces a systematic bias in image brightness. In particular, this fixed ordering enforces $\mathbb{E}[x_{1}] \leq \mathbb{E}[x_{2}]$, which violates the fundamental assumption of Noise2Noise (N2N). To address this mismatch, we further introduce randomness. For each spatial location $(r, c)$ in the sampled pairs, we draw a random variable $p \sim \mathcal{U}[0, 1)$ and assign the values as
\begin{equation}
(x_1,\ x_2)_{r,c} = 
\begin{cases}
(m_1,\ m_2), & \text{if } p < 0.5, \\
(m_2,\ m_1), & \text{otherwise}.
\end{cases}
\label{eq:random2}
\end{equation}
This randomization breaks the fixed ordering of $m_1 \leq m_2$ and ensures that $x_1$ and $x_2$ have, in expectation, the same mean intensity. As a result, the paired samples better satisfy the independence assumption required for effective N2N training.

Furthermore, the randomized assignment process in Eqs.~\ref{eq:random1} and \ref{eq:random2} provides three distinct sampling outcomes for each pixel pair, calculated as
\[
\binom{3}{2} \times \binom{2}{1} \times \frac{1}{2},
\]
where the factor $\frac{1}{2}$ arises from the symmetric loss that will be introduced in the next section. Assuming that $\tilde{I}$ is partitioned into $N$ patches, the sampling space associated with each sampling position is of size $3^N$.

\subsection{Noise2Noise Training}
Unlike the intuitive Noise2Clean (N2C) paradigm, N2N trains a denoising network $f_\theta$ to predict one noisy observation from another. It assumes that the two inputs, $x_1$ and $x_2$, are noisy observations of the same clean image corrupted by independent zero-mean noise. Under this assumption, the N2N objective is defined as
\begin{equation}
\mathcal{L}_{\text{N2N}} = \mathbb{E}_{(x_1, x_2) \sim X} \left[ \| f_\theta(x_1) - x_2 \|_2^2 \right],
\end{equation}
where $X$ denotes a dataset of paired noisy images. 

Through the M2M sampling process described in Subsection~\ref{sec:sampling algorithm}, pseudo-independent noisy pairs are constructed from a single image $I$ corrupted by structured noise. These sub-image pairs are then used for N2N training. Specifically, in each complete M2M sampling instance, the nine sampling positions yield nine sub-image pairs fully covering the original noisy image, which are used to train nine independent networks of the same architecture. The training objective of these networks incorporates two extensions to the standard N2N formulation: a symmetric loss~\cite{chen2021exploring} and a consistency loss~\cite{mansour2023zero}.

First, to enhance training stability and better exploit the mutual information within each pseudo pair, we employ a symmetric loss in which the two elements of a pair alternately serve as source and target:
\begin{equation}
\mathcal{L}_{\text{sym}} = \frac{1}{2} \mathbb{E}_{(x_1, x_2)\sim I} \left[ \| f_\theta(x_1) - x_2 \|_2^2 + \| f_\theta(x_2) - x_1 \|_2^2 \right],
\label{eq:symloss}
\end{equation}
where $(x_1, x_2)$ denote a noisy sub-image pair obtained by M2M sampling at one of the nine sampling positions across all patches of the original noisy image $I$.

Second, to enforce agreement between sampling and denoising operations, we extend the consistency loss to structured-noise settings. The core idea is that denoising followed by sampling should yield results consistent with sampling followed by denoising. Unlike the original design for i.i.d. noise, our formulation explicitly incorporates structured noise by first de-structuring the input. Specifically, the noisy image is processed through M2M sampling at nine sampling positions across all patches, each yielding a sub-image pair, which is averaged to produce a single sub-image. The nine resulting sub-images are then reassembled according to their sampling positions to reconstruct a de-structured image $\hat{I}$ with the same dimensions as the original input $I$. The above de-structuring process ensures that the network consistently operates on inputs of a uniform type. The consistency loss is then defined as
\begin{equation}
\mathcal{L}_{\text{cons}}(\theta) = \tfrac{1}{2}\,\mathbb{E}_{(x_1, x_2)\sim I}\!\left[ \| \hat{y}_1 - f_\theta(x_1) \|_2^2 + \| \hat{y}_2 - f_\theta(x_2) \|_2^2 \right],
\end{equation}
where $(x_1, x_2)$ follows the same definition as in Eq.~\ref{eq:symloss}, and $(\hat{y}_1, \hat{y}_2)$ denotes the corresponding denoised sub-image pair obtained from the same sampling position across all patches of $f_\theta(\hat{I})$.

Finally, the total objective combines the symmetric and consistency terms:
\begin{equation}
\mathcal{L}_{\text{total}}(\theta) = \mathcal{L}_{\text{sym}}(\theta) + \lambda\,\mathcal{L}_{\text{cons}}(\theta),
\end{equation}
where $\lambda$ balances their contributions.

\begin{figure}[t]
\centering
\includegraphics[width=0.9\linewidth]{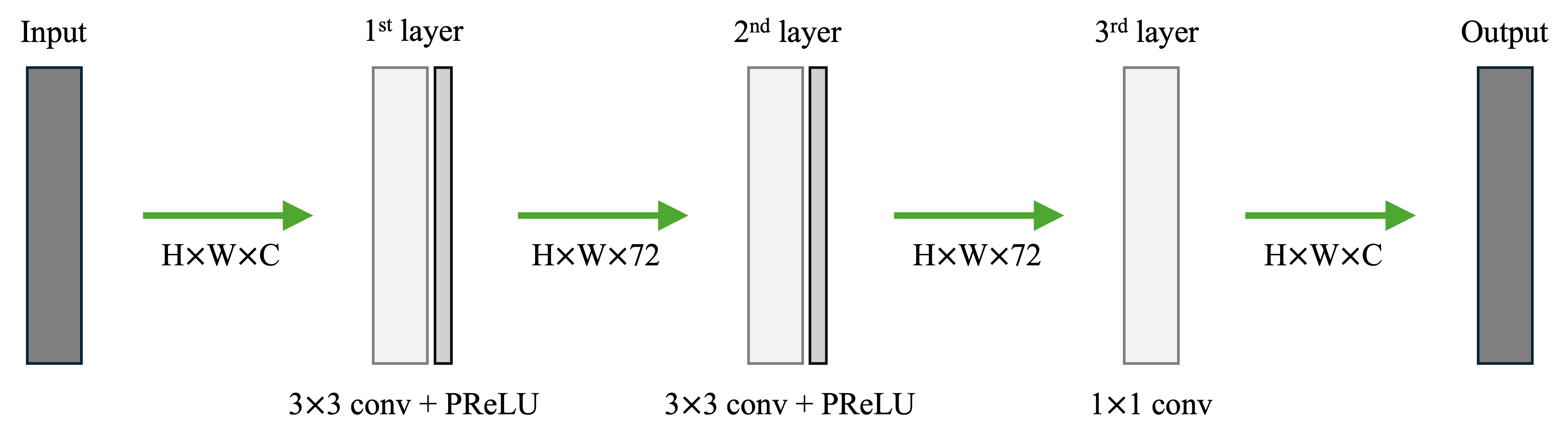}
\caption{Three-layer convolutional neural network denoiser. The first two layers employ $3\times 3$ convolutions followed by PReLU, while the final layer is a $1\times 1$ convolution that produces the denoised output. The image dimensions at each stage are indicated below the arrows.}
\label{fig:network_architecture}
\end{figure}

\subsection{Network Architecture}
Since zero-shot learning does not rely on external training data, using overly deep networks could overfit \cite{lequyer2022fast}. To ensure stability and generalizability, we adopt a lightweight three-layer convolutional neural network as the denoiser, as illustrated in Fig.~\ref{fig:network_architecture}. The first two layers use $3 \times 3$ convolutional kernels with a stride of 1, each followed by a PReLU activation function. Each convolutional layer embeds the input into 72 feature channels to capture rich representations while keeping the model compact. The third layer is a $1 \times 1$ convolution with stride 1, serving as the output layer that reconstructs the denoised image. The network operates on inputs that have been de-structured via M2M filtration. For training, we instantiate nine independent networks, each of which is associated with a predefined sampling position. These networks share the same architecture but are trained in parallel on their corresponding sub-image pairs.

\section{Simulated Experiments}

In this section, we evaluated the denoising performance of our proposed method, Median2Median (M2M). For comparison, we consider several state-of-the-art blind zero-shot baselines, namely Noise2Fast (N2F) \cite{lequyer2022fast} and Zero-Shot Noise2Noise (ZS-N2N) \cite{mansour2023zero}, as well as the classical non-blind method BM3D \cite{dabov2007image}. All experiments were simulated on grayscale images corrupted by synthetic correlated noise. We employed two public datasets: the Kodak dataset \cite{kodak} and a fluorescence microscopy dataset \cite{zhang2019poisson}. Denoising performance was measured using two standard metrics, Peak Signal-to-Noise Ratio (PSNR) and Structural Similarity Index (SSIM) \cite{wang2004image}.

\subsection{Synthetic Structured Noise}

Structured noise is commonly observed in real-world imaging scenarios, such as CT and fluorescence microscopic images. These noise patterns often appear as banding, striping, or other directional artifacts with strong local correlations along a preferred direction. To simulate structured noise in grayscale images, we first generate a 2D independent and identically distributed (i.d.d.) noise field $x \in \mathbb{R}^{H \times W}$:
\begin{equation}
x \sim \mathcal{N}(0, \sigma_0^2 I),
\end{equation}
where $I$ is the identity matrix, and $\sigma_0^2$ denotes the initial variance. To introduce correlation along an arbitrary angle $\theta$, we convolve the noise image with a directional averaging kernel:
\begin{equation}
y = x \ast h_\theta,
\end{equation}
where
\begin{equation}
h_\theta = \Big[\tfrac{1}{\ell}, \tfrac{1}{\ell}, \dots, \tfrac{1}{\ell}\Big] \in \mathbb{R}^{1 \times \ell}
\end{equation}
is a one-dimensional averaging kernel of length $\ell$, and $\ell$ is an odd integer to give a well-defined center. More explicitly, let $(p,q)\in\mathbb{Z}^2$ be the smallest integer offset vector determined by the given angle $\theta$, and let the geometric step length be $s=\sqrt{p^2+q^2}$. To avoid redundant representations, we restrict the angle to $\theta \in [0,\pi)$, and equivalently allow $p$ to take any integer while $q$ is restricted to positive integers. For example, $(1,1)$ corresponds exactly to $\theta=45^\circ$, while $(-1,2)$ corresponds to $\theta=\pi-\arctan(2)\approx 116.6^\circ$. 
For simplicity of notation, in the following we will use $\theta$ to denote the direction determined by $(p,q)$. 
The directional averaging along $\theta$ is defined as
\begin{equation}
y(i,j) = \frac{1}{\ell}\sum_{k=-r}^{r} x\!\big(i+k\,p,\ j+k\,q\big), \quad r=\frac{\ell-1}{2},
\end{equation}
with symmetric padding applied at the image boundaries. This operation introduces directional correlation along $\theta$, where the spacing between adjacent samples equals $s$.
The variance of the filtered noise becomes
\begin{equation}
\mathrm{Var}(y(i,j)) = \frac{1}{\ell}\sigma_0^2,
\end{equation}
since $y(i,j)$ is the mean of $\ell$ i.i.d.\ Gaussian samples taken at distinct grid points.
The covariance between pixels separated by $t$ discrete steps along $\theta$ (i.e., geometric distance $d=t\,s$) is
\begin{equation}
\mathrm{Cov}\big(y(i,j),\, y(i+t\,p,\, j+t\,q)\big) 
= 
\begin{cases}
\frac{\ell - |t|}{\ell^2}\,\sigma_0^2, & |t| < \ell, \\[6pt]
0, & |t| \geq \ell ,
\end{cases}
\end{equation}
while the covariance between pixels separated orthogonally to $\theta$ (i.e., along a perpendicular direction $(-q,p)$) remains
\begin{equation}
\mathrm{Cov}\big(y(i,j),\, y(i+t(-q),\, j+t\,p)\big) = 0.
\end{equation}
Accordingly, the correlation coefficient between pixels separated by $t$ steps along $\theta$ is
\begin{equation}
\rho_\theta(t) 
= \frac{\mathrm{Cov}(y(i,j),\, y(i+t\,p,\, j+t\,q))}
{\mathrm{Var}(y(i,j))}
= 
\begin{cases}
\frac{\ell - |t|}{\ell}, & |t| < \ell, \\[6pt]
0, & |t| \geq \ell .
\end{cases}
\end{equation}
The above derivations indicate that $\sigma_0$ influences the noise level, while the window length $\ell$ determines the correlative structure.

To match a target noise level $\sigma_n$ that directly specifies the strength of synthetic noise, the filtered noise can be scaled as follows:
\begin{equation}
\tilde{y}(i,j) = \sigma_n \cdot \frac{y(i,j)}{\mathrm{std}(y)},
\end{equation}
where $\mathrm{std}(y)$ denotes the empirical standard deviation of the filtered noise $y$. Finally, a correlated noise $\tilde{y}$ is added to a clean image $I_{\text{clean}}$ to obtain a noisy observation:
\begin{equation}
I_{\text{noisy}} = \mathrm{clip}\left(I_{\text{clean}} + \tilde{y},\, 0,\, 1\right),
\end{equation}
where $\mathrm{clip}(\cdot)$ constrains the pixel intensities to the range $[0,1]$.

\subsection{Comparison on Benchmarks}

\begin{figure}[t]
    \centering
    \includegraphics[width=1\linewidth]{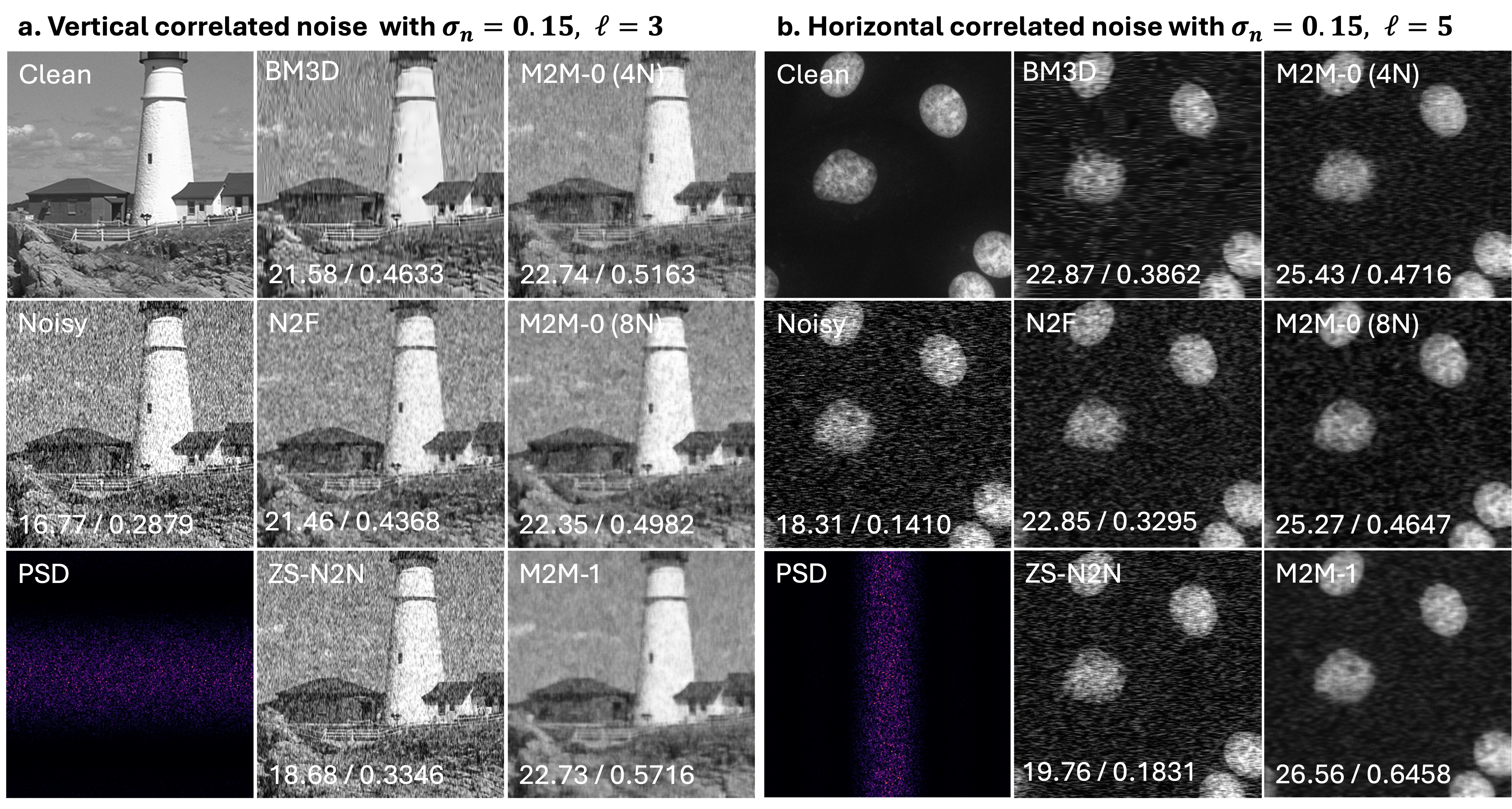}
    \caption{Visual comparison of representative results obtained for synthetic one-dimensionally correlated noise. 
    (a) A Kodak image with vertical correlated noise ($\sigma_n = 0.15$, $\ell = 3$).
    (b) A fluorescence microscopic image with horizontal correlated noise ($\sigma_n = 0.15$, $\ell = 5$). 
    For M2M, “M2M-1” denotes the version using first-order interpolation, whereas “M2M-0 (4N)” and “M2M-0 (8N)” denote the versions using zero-order interpolation with four- and eight-neighbors, respectively. The power spectral density (PSD) of the synthetic noise is shown in the bottom-left corner of each panel, highlighting the directional correlation introduced by structured noise. Quantitative results are reported as PSNR (dB)/SSIM.}
    \label{fig:visual_comparison}
\end{figure}

We first conducted a visual comparison between the proposed M2M method and state-of-the-art zero-shot denoising approaches, including N2F, ZS-N2N, and the classical BM3D. Fig.~\ref{fig:visual_comparison} shows the results on images corrupted by synthetic vertical and horizontal structured noise. Although the noisy images exhibit strong local correlations, the denoising results show that M2M effectively removed dominant structured components from the noisy images while preserving meaningful details. Among the M2M variants, the zero-order interpolation methods tend to preserve more fine details, whereas the first-order interpolation method achieves stronger suppression of structured noise. In contrast, the other zero-shot methods failed to adequately suppress the structured noise, leaving most of the correlated artifacts essentially intact.

\begin{table*}[t]
\centering
\caption{Performance comparison among the zero-shot denoising methods on fluorescence microscopic images corrupted by synthetic horizontal structured noise. Results are reported as PSNR (dB) / SSIM. $\ell$ denotes the kernel length of the averaging filter and $\sigma_n$ represents the target noise level (standard deviation). For $\ell=1$, the structured noise degenerates to i.d.d. Gaussian noise. The best results for each setting are highlighted in bold.}
\label{tab:zs_denoising_comparison}
\renewcommand{\arraystretch}{1.5}
\rowcolors{2}{white}{darkmossgreen!30} 
\begin{tabularx}{\textwidth}{
  >{\centering\arraybackslash}m{0.3cm}     
  >{\centering\arraybackslash}m{0.5cm}     
  >{\centering\arraybackslash}m{2cm}       
  >{\centering\arraybackslash}m{2cm}       
  >{\centering\arraybackslash}X            
  >{\centering\arraybackslash}m{2cm}       
  >{\centering\arraybackslash}m{2cm}       
  >{\centering\arraybackslash}m{2cm}}      
\Xhline{0.8pt}
\rowcolor{darkmossgreen!90} $\boldsymbol{\ell}$ & $\boldsymbol{\sigma_n}$ & \textbf{BM3D~\cite{dabov2007image}} & \textbf{Noise2Fast~\cite{lequyer2022fast}} & \textbf{Noise2Noise\textsuperscript{*}~\cite{mansour2023zero}} & \multicolumn{3}{c}{\textbf{Median2Median}} \\ \rowcolor{darkmossgreen!90} & & & & & \textbf{0-order (4N)} & \textbf{0-order (8N)} & \textbf{1-order} \\
\hline
\textbf{1} & 0.05 & \textbf{37.14} / \textbf{0.9111} & 36.08 / 0.8949 & 35.29 / 0.8748 & 35.46 / 0.8855 & 35.58 / 0.8877 & 35.67 / 0.8904 \\
           & 0.10 & \textbf{32.78} / \textbf{0.8242} & 31.23 / 0.7962 & 30.80 / 0.7711 & 32.31 / 0.8009 & \textbf{32.74} / 0.8027 & 32.35 / 0.8116 \\
           & 0.15 & 29.29 / \textbf{0.7513} & 27.54 / 0.7106 & 27.28 / 0.6799 & 29.68 / 0.7231 & \textbf{30.54} / 0.7154 & 29.27 / 0.7364 \\
\textbf{3} & 0.05 & 32.39 / 0.7670 & 31.64 / 0.7174 & 28.99 / 0.5668 & 33.73 / 0.8289 & 33.46 / 0.8217 & \textbf{34.30} / \textbf{0.8607} \\
           & 0.10 & 28.21 / 0.6082 & 27.10 / 0.4983 & 24.18 / 0.3194 & 29.61 / 0.6587 & 29.43 / 0.6495 & \textbf{30.25} / \textbf{0.7314} \\
           & 0.15 & 25.66 / 0.5197 & 23.89 / 0.3456 & 21.07 / 0.1935 & 26.68 / 0.5221 & 26.75 / 0.5137 & \textbf{27.06} / \textbf{0.6135} \\
\textbf{5} & 0.05 & 30.19 / 0.6577 & 30.74 / 0.6835 & 27.88 / 0.5161 & 32.34 / 0.7712 & 32.37 / 0.7835 & \textbf{33.47} / \textbf{0.8393} \\
           & 0.10 & 25.66 / 0.4548 & 25.96 / 0.4476 & 22.91 / 0.2745 & 27.87 / 0.5572 & 28.02 / 0.5864 & \textbf{29.25} / \textbf{0.6879} \\
           & 0.15 & 23.00 / 0.3512 & 22.77 / 0.3022 & 19.75 / 0.1616 & 24.97 / 0.4116 & 25.23 / 0.4461 & \textbf{26.12} / \textbf{0.5594} \\
\textbf{7} & 0.05 & 29.23 / 0.6062 & 30.29 / 0.6669 & 27.44 / 0.4989 & 31.54 / 0.7334 & 31.79 / 0.7612 & \textbf{32.98} / \textbf{0.8245} \\
           & 0.10 & 24.53 / 0.3896 & 25.33 / 0.4240 & 22.26 / 0.2549 & 26.92 / 0.5031 & 27.24 / 0.5507 & \textbf{28.69} / \textbf{0.6618} \\
           & 0.15 & 21.80 / 0.2894 & 22.34 / 0.2909 & 19.26 / 0.1548 & 24.07 / 0.3621 & 24.40 / 0.4120 & \textbf{25.61} / \textbf{0.5276} \\
\Xhline{0.8pt}
\end{tabularx}
\vspace{1.5em}
\raggedright\footnotesize * Noise2Noise refers to Zero-shot Noise2Noise (ZS-N2N).
\par
\end{table*}

Table~\ref{tab:zs_denoising_comparison} reports the quantitative evaluation of these zero-shot denoising methods on fluorescence microscopic images corrupted with synthetic structured noise along the horizontal direction. The experiments consider both i.i.d. Gaussian noise and structured correlated noise, parameterized by the kernel length $\ell$ and noise variance $\sigma_n$. When $\ell > 1$, the structured noise produces correlated noise patterns, whereas for $\ell = 1$, it degenerates into i.i.d. Gaussian noise. Since BM3D is a non-blind method, the true noise variance $\sigma_n$ was provided as input. 
In the i.i.d. case ($\ell = 1$), M2M remains competitive with state-of-the-art zero-shot methods. At a low noise level, M2M is slightly below BM3D and Noise2Fast, but as the noise strength increases, it consistently achieves the best results among blind zero-shot methods—and in several cases, even surpasses the non-blind BM3D baseline.
For $\ell > 1$, the noise exhibits spatial correlations and becomes substantially more challenging. In these settings, all other zero-shot methods rapidly lose effectiveness. In contrast, M2M consistently suppresses the correlated noise and achieves the best denoising performance among all zero-shot methods across all tested correlation lengths and noise levels.

Overall, these findings demonstrate that M2M is robust and effective across both independent and correlated noise conditions. It maintains competitiveness in the simpler i.i.d. case and shows clear superiority in the more challenging correlated settings, establishing itself as the only blind zero-shot method that consistently delivers state-of-the-art performance across both regimes. Moreover, although M2M deploys nine independent lightweight networks, it remains relatively efficient: for noisy images of the same size, its denoising time is only about six times that of N2F or ZS-N2N, which is acceptable in practical applications.

\subsection{Ablation Study}
Beyond the core idea of directional interpolation and generalized median filtration, several additional techniques are essential for M2M to effectively suppress structured noise. These include block-wise sampling, the inclusion of the target pixel in the candidate set, the random assignment (RA), and the repeated inference (RI).

\begin{figure}[t]
\centering
\includegraphics[width=\linewidth]{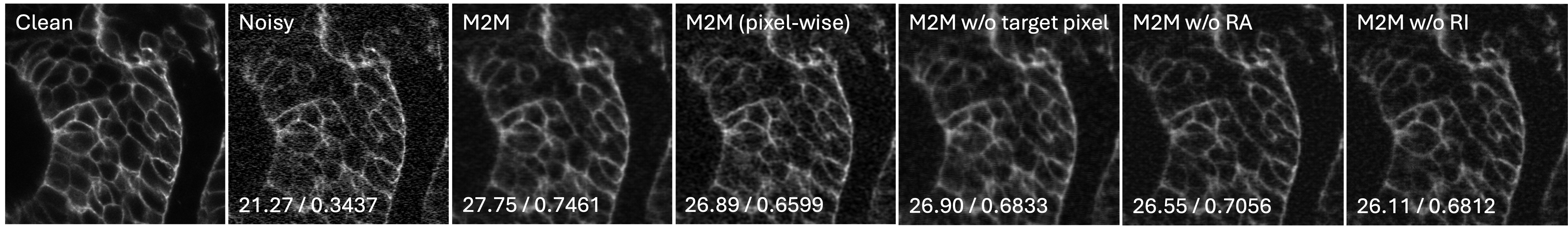}
\caption{Ablation study on each key ingredient of M2M for a fluorescence microscopic image corrupted with synthetic structured noise ($\sigma_n = 0.1$, $\ell = 3$). The first two images show the clean reference and the corresponding noisy input. The subsequent results are obtained with (i) the full-fledged M2M using first-order interpolation, and from this baseline, (ii) removing block-wise sampling (replaced with pixel-wise sampling), (iii) removing the center pixel from the candidate set, (iv) removing the randomized assignment strategy, and (v) removing the repeated inference process.}
\label{fig:ablation}
\end{figure}

To validate the necessity of these design ingredients, we conducted an ablation study on a fluorescence microscopic image corrupted with synthetic structured noise ($\sigma_n = 0.1$, $\ell = 3$). Fig.~\ref{fig:ablation} illustrates the results when each component was individually removed from the M2M version with first-order interpolation. All ablation variants substantially compromised the denoising performance compared to the full-fledged M2M, confirming that each design choice is indispensable for achieving optimal outcomes.

\section{Discussions and Conclusion}

Beyond 2D images, the M2M framework can be naturally generalized to higher-dimensional data. For any data tensor, directional interpolation can be applied to its sub-tensors along multiple orientations, producing a collection of candidate signal estimates analogous to those in 2D M2M. These sub-tensors provide richer contextual information across spatial, temporal, or spectral dimensions, allowing the formation of pseudo-independent noisy pairs even in highly structured noise environments. We hypothesize that this increased dimensionality will further enhance the denoising performance, as more diverse and complementary directional estimates become available for robust median-based selection and Noise2Noise training.

In this feasibility study, we explored zero-order and first-order interpolation for directional estimation, selected for their simplicity and computational efficiency. Zero-order interpolation with nearest-neighbor approximation helps preserve sharp edges by minimizing blurring effects, whereas first-order interpolation with linear approximation produces smoother results. Nevertheless, higher-order methods such as cubic interpolation merit further exploration, as they have the potential to provide more accurate intensity estimation by incorporating richer local information. A systematic investigation of these strategies could reveal trade-offs between efficiency, edge preservation, and denoising performance, thereby enabling tailored solutions for specific imaging modalities, noise characteristics, and real-world applications.

Another promising direction is to integrate the M2M framework with the non-local means (NLM) concept for denoising~\cite{niu2022noise}. Instead of relying solely on local directional interpolation within a single sub-tensor, we can search for and incorporate similar sub-tensors across the entire image or volume. These non-local sub-tensors provide additional candidate signal estimates for each target pixel or voxel, thereby expanding the pool of pseudo-independent observations. By combining M2M’s median-awared selection with ensemble learning across multiple sub-tensor estimates, we may further improve the robustness and accuracy of denoising, especially in scenarios with repetitive structures or textures.

In conclusion, we have introduced M2M, a novel zero-shot denoising framework for suppressing structured noise in images. Unlike existing zero-shot approaches whose effectiveness is limited to i.i.d. noise, M2M generates pseudo-independent sub-image pairs from images corrupted by structured noise through directional interpolation and generalized median filtration, adaptively avoiding values distorted by structured artifacts. Our randomized assignment strategy further enlarges the effective sampling space and eliminates systematic bias, ensuring that the sampled pairs remain suitable for Noise2Noise training. Consequently, M2M provides an efficient and training data-free solution for handling structured noise. Experimental results demonstrate that M2M is comparable to state-of-the-art zero-shot denoising methods in the i.i.d. case, while consistently outperforming them in the correlated case. This work takes the first step toward zero-shot denoising of structured noise, opening a door for effective noise removal from a single noisy image without the strict i.i.d. assumption.

\section*{Acknowledgments}
This project is funded by the National Institute of Health R01 CA233888. The content is solely the responsibility of
the authors and does not necessarily represent the official views of the National Institutes of Health. 

\bibliographystyle{unsrt}  
\bibliography{main}

\end{document}